# LADDER AND SUBDIVISION OF LADDER GRAPHS WITH PENDANT EDGES ARE ODD GRACEFUL

M. I. Moussa
Computer Science Department, Faculty of Computers and Informatics, Benha University, Benha, Egypt.

E.M. Badr
Scientific Computing Department, Faculty of Computers and Informatics, Benha University, Benha, Egypt.

*Abstract:*

*The ladder graph plays an important role in many applications as Electronics, Electrical and Wireless communication areas. The aim of this work is to present a new class of odd graceful labeling for the ladder graph. In particular, we show that the ladder graph $L_n$ with m-pendant $L_n \odot mk_1$ is odd graceful. We also show that the subdivision of ladder graph $L_n$ with m-pendant $S(L_n) \odot mk_1$ is odd graceful. Finally, we prove that all the subdivision of triangular snakes ($\Delta_k - snake$) with pendant edges $S(\Delta_k - snake) \odot mk_1$ are odd graceful.*

INTRODUCTION:

A labeling of the graph is assigning a nature values to the vertices of the graph in some way that induced edge labels according to certain pattern. Given a graph $G$ ($V$, $E$) with $p$ vertices and $q$ edges. A graceful labeling is an injection $\phi : V(G) \to \{0,1,2,...,q\}$ such that the induced function given by $\phi^*(u,v) = |\phi(u) - \phi(v)|$ for all edge ($u$,$v$). Interest in graph labeling began in 1967 when Rosa published his paper on graph labeling, he called a function $\phi$ a β-labeling of a graph G subsequently called such labeling graceful [1]. Many types of graph labeling were introduced for example, magic total, ant-magic total, cordial, harmonic, graceful and odd graceful. The definition of the graph odd graceful was proposed by Gnanajothi (1991) [2] as follow: a graph $G$ of size $q$ is odd-graceful if there is an injection $\phi$ from $V(G)$ to $\{0, 1, 2, …, 2q-1\}$ such that, when each edge $xy$ is assigned the label or weight $|\phi(x) - \phi(y)|$, the resulting edge labels are $\{1, 3, 5, …, 2q-1\}$.

**Defintion1** : A Cartesian product of two subgraph $G_1$ and $G_2$ is the graph $G_1 \times G_2$ such that its vertex set is a Cartesian product of $V(G_1)$ and $V(G_2)$ i.e. $V(G_1 \times G_2) = V(G_1) \times V(G_2) = \{ (x, y): x$ $\in V(G_1)$, $y \in V(G_2)$ } and its edge set is defined as $E(G_1 \times G_2) = \{ ((x_1,x_2),(y_1,y_2)) : x_1 = y_1$ and $(x_2,y_2) \in E(G_2)$ or $x_2 = y_2$ and $(x_1,y_1) \in E(G_1)$ }.

**Defintion2 :** The ladder graph $L_n$ is defined by $L_n = P_n \times K_2$ where $P_n$ is a path with $n$ vertices and x denotes the Cartesian product and $K_2$ is a complete graph with two-vertices.

Several classes of the ladder graph labelings have been discovered. A first family of methods were introduced within the ladder graph labeling field is based on the graceful labeling. That the ladder graph $P_m \times P_2$ are graceful was proved by Maheo (1980) [3]. More general approach was given by Acharya and Gill (1981) [4] for the two-dimensional square lattice graphs $P_m \times P_n$. A second family of methods are based on harmonious labeling of ladder graphs. The ladder graphs $P_m \times P_2$ are harmonious when $m > 2$ was proved by Graham and Sloane(1980)[5]. Kathiresan has proved that the ladder graphs are graceful labeling [6]. A slightly different scheme is based on subdivision ladder graph within this scheme, the graphs obtained from the ladder by subdividing each edge exactly once are graceful [6,7]. The odd-Graceful labeling of subdivision of ladder graphs was introduced by Badr et al [8].

In this paper, we show that the ladder graph $L_n$ with $m$-pendant $L_n \square mk_1$ is odd graceful. We also show that the subdivision of ladder graph $L_n$ with $m$-pendant $S(L_n) \square mk_1$ is odd graceful. Finally, we prove that all the subdivision of triangular snakes ($\Delta_k - snake$) with pendant edges $S(\Delta_k - snake) \square mk_1$ are odd graceful.

## 2. APPLICATIONS OF LADDER GRAPH

In this section, we introduce three applications which related to the ladder graph. Because of the previous three applications and many others, we motivate to study the odd graceful labeling for the ladder graph.

The first application is electronics filed. Resistor ladder networks provide a simple, inexpensive way to perform digital to analog conversion (DAC). The most popular networks are the binary weighted ladder and the R/2R ladder. Both devices will convert digital voltage information to analog, but the R/2R ladder has become the most popular due to the network's inherent accuracy superiority and ease of manufacture.

The second application is electrical area. The flow graph of a ladder is developed using Ohm's and the two Kirchhoff's equations. The graph is then reciprocated so that it contains only forward paths. The transfer function, in the case of a simple ladder, is the reciprocal of the sum of all distinct paths from the output to the input node. In the case of ladders containing internal generators, independent or dependent, the transfer function can be found in a similar way with slight modifications. Other relations, such as the input impedance, transfer admittance, etc., can also be found directly from the flow graph.

The third application is wireless communication area. Throughout time, more and more wireless networks have been developed to ease communication between any two systems (computers, telephones, etc.) without the use of wires. But the available radio frequencies allocated to wireless communications are insufficient (In the U.S. all WiFi communications for the 2.4 GHz band are limited to 11 channels). It is important to find an efficient way to have safe transmissions in areas such as Cellular telephony, WiFi, Security systems and many more [9]. It is unpleasant being on the phone and getting someone else on the same line. This inconvenience is given by interferences caused by unconstrained simultaneous transmissions [10]. Two close enough channels can interfere or resonate thereby damaging communications. The interference can be avoided by means of a suitable channel assignment. In order to found the solution, Hale [11] formulated the problem into the notion of a graph (vertex) coloring model which we call the $L(2, 1)$-coloring today. Suppose a number of transmitters or stations are given. We have to assign a channel to each of the given transmitters or station such that the interference is minimized. If the transmitters in physical space or the frequency channels they use are too close to each other, the interference phenomena may degrade the quality of time sensitive or error sensitive communication. In order to minimize the interference, any two "close" transmitters must have maximally different channels assigned.

## 3. MAIN RESULTS

**Theorem1:** The ladder graph $L_n$ with *m-pendant* edges $L_n \odot mk_1$ is odd graceful.

**Proof:**

Let $G = L_n \odot mk_1$ has $q$ edges and $p$ vertices. The graph $G$ consists of the vertices $u_1u_2\ldots u_n$, $v_1v_2\ldots v_n$ which are the vertices of the ladder graph $L_n$, each of $u_i$ connect with *m-pendant* edges $u_i^j$ and each of $v_i$ connect with *m-pendant* edges $v_i^j$ where $i = 1, 2, \ldots, n$ and $j = 1, 2, \ldots, m$. Clearly, the graph $G = L_n \odot mk_1$ has $q = 2mn + 3n - 2$ edges and $p = 2n(m+1)$ vertices.

We prove that the ladder graph $L_n$ with *m-pendant* $L_n \odot mk_1$ is odd graceful. Let us consider the following numbering $\phi$ of the vertices of the graph $G$:

$\phi(v_i) = i - 1$                      , $i$ odd: $1 \leq i \leq n$

$\phi(v_i) = 2q - i + 1$                , $i$ even: $1 \leq i \leq n$

$\phi(u_i) = 2q - i - 2n + 2$           , $i$ odd: $1 \leq i \leq n$

$\phi(u_i) = 2n + i - 2$               , $i$ even: $1 \leq i \leq n$

$\phi(v_i^j) = 2m(i-1) + 2j + 1$         , $i$ odd: $3 \leq i \leq n$, $1 \leq j \leq m$

$\phi(v_i^j) = 2q - (2m + 1)i - 2j + 2m + 2$    , $i$ even: $2 \leq i \leq n$, $1 \leq j \leq m$

$\phi(v_1^j) = 2j - 1$                     $1 \leq j \leq m$

$\phi(u_i^j) = 2q - (2m + 1)i - 2j - (2m + 2)n + 2m + 3$    , $i$ odd: $1 \leq i \leq n$, $1 \leq j \leq m$

$\phi(u_i^j) = 2q + (2m + 1)i + 2j - (2m + 4)n - 2m + 1$    , $i$ even: $1 \leq i \leq n$, $1 \leq j \leq m$

a)
$$Max_{v \in V}\phi(v) = \max\left\{\max_{\substack{i\text{ odd}\\1\leq i\leq n}}(i-1), \max_{\substack{i\text{ even}\\1\leq i\leq n}}(2q-i+1), \max_{\substack{i\text{ odd}\\1\leq i\leq n}}(2q-i-2n+2), \max_{\substack{i\text{ even}\\1\leq i\leq n}}(2n+i-2), \max_{\substack{1\leq j\leq m\\1\leq i\leq n\\i\text{ odd}}} 2m(i-1)+2j+1,\right.$$

$$\max_{\substack{1\leq j\leq m\\1\leq i\leq n\\i\text{ even}}} 2q-(2m+1)i-2j+2m+2, \max_{\substack{1\leq j\leq m}} 2j-1, \max_{\substack{1\leq j\leq m\\1\leq i\leq n\\i\text{ odd}}} 2q-(2m+1)i-2j-(2m+2)n+2m+3,$$

$$\left.\max_{\substack{1\leq j\leq m\\1\leq i\leq n\\i\text{ odd}}} 2q+(2m+1)i+2j-(2m+4)n-2m+1\right\} = 2q-1$$

, the maximum value of all odd integers. Thus $\phi(v) \in \{0, 1, 2 \ldots, 2q-1\}$

(b) Clearly $\phi$ is a one – to – one mapping from the vertex set of $G$ to $\{0, 1, 2, \ldots, 2q-1\}$.

c) It remains to show that the labels of the edges of $G$ are all the odds of the interval $[1, 2q-1]$.

The range of $|\phi(v_j) - \phi(v_i)| = \{2q - j - i + 2$    , $j$-even, $i$-odd, $1 \leq i, j \leq n\}$

The range of $|\phi(u_j) - \phi(u_i)| = \{2q - j - i - 4n + 4$    , $j$-odd, $i$-even, $1 \leq i, j \leq n\}$

The range of $|\phi(v_1^j) - \phi(v_1)| = \{2j - 1, 1 \leq j \leq m\}$

The range of $|\phi(v_i^j) - \phi(v_i)| = \{2m(i-1) + 2j - i + 2, i \text{ odd}, 3 \leq i \leq n, 1 \leq j \leq m\}$

The range of $|\phi(v_i^j) - \phi(v_i)| = \{2mi + 2j - 2m - 1, i \text{ even}, 1 \leq i \leq n, 1 \leq j \leq m\}$

The range of $|\phi(u_i^j) - \phi(u_i)| = \{2mi - 2j - 2mn - 2m - 1, i \text{ odd}, 1 \leq i \leq n, 1 \leq j \leq m\}$

The range of $|\phi(u_i^j) - \phi(u_i)| = \{2q + 2mi + 2j - 2m(n+1) - 3(2n-1), i \text{ even}, 1 \leq i \leq n, 1 \leq j \leq m\}$

Hence $\{|\phi(u) - \phi(v)| : uv \in E\} = \{1, 3, \ldots, 2q-1\}$ so that *graph* $G = L_n \odot mk_1$ is odd graceful.

**Theorem 2:** The subdivision of ladder graph $L_n$ with *m-pendant* edges $S(L_n) \odot mk_1$ is odd graceful.

**Proof:**

We can see a subdivision of ladder graph $S(L_n)$ as the graph which consists of two paths, a right path $v_1 v_2 .. v_{2n-1}$ and a left path $u_1 u_2 .. u_{2n-1}$. In order to get the subdivision of ladder graph $S(L_n)$, we run the following steps:

i) Connect the vertex $u_i$ with $v_i$ where i = 1, 2, …, 2n-1.

ii) Let $w_i$ be the newly added vertex between $u_i$ and $v_i$ where i= 1, 3, 5,…, 2n-1.

Now, in order to get the subdivision of ladder graph $S(L_n)$ with pendant edges $S(L_n) \odot mk_1$, for each $u_i$, $v_j$, $w_k$ connect with m-pendant edges where i = 1, 2, …, 2n-1, j = 1, 2, …, 2n-1 and k = 1, 3, 5, …, 2n-1. Clearly, the graph $G = L_n \odot mk_1$ has $q = m(5n-2)+2(3n-2)$ edges and $p = (5n-2)(m+1)$ vertices.

We prove that the ladder graph $L_n$ with *m-pendant* $S(L_n) \odot mk_1$ is odd graceful. Let us consider the following numbering $\phi$ of the vertices of the graph *G*:

$\phi(v_i) = i - 1$ , $i$ – odd, $1 \leq i \leq 2n-1$.
$\phi(v_i) = 2q - i + 1$ , $i$ – even, $2 \leq i \leq 2n-2$
$\phi(w_i) = 2q - 1 - 4n + 4(i-1)$ $1 \leq i \leq n$
$\phi(u_i) = i + 2n - 1$ , $i$ –odd, $1 \leq i \leq 2n-1$
$\phi(u_i) = 2q - i - 6n + 5$ , $i$ –even, $2 \leq i \leq 2n-2$
$\phi(v_i^j) = (2m+1)i + 2j - 2m - 2$ , $i$ –odd, $1 \leq i \leq 2n-1, 1 \leq j \leq m$
$\phi(v_i^j) = 2q - (2m+1)i - 2j + 2m + 2$ , $i$ – even, $2 \leq i \leq 2n-2, 1 \leq j \leq m$
$\phi(w_i^j) = 2q + (2m+4)i + 2j - (6m+6)n$ $1 \leq i \leq n, 1 \leq j \leq m$
$\phi(u_i^j) = 2q + (2m+1)i + 2j - (4m+10)n + 6$ , $i$ –odd, $1 \leq i \leq 2n-1, 1 \leq j \leq m$
$\phi(u_i^j) = q - (2m+1)i - 2j - mn + 2m + 2$ , $i$ –even, $2 \leq i \leq 2n-2, 1 \leq j \leq m$

$$Max_{v \in V} \phi(v) \max \left\{ \max_{\substack{1 \leq i \leq 2n-1 \\ i \text{ odd}}} (i-1), \max_{\substack{2 \leq i \leq 2n-2 \\ i \text{ even}}} (2q-i+1), \max_{\substack{1 \leq i \leq 2n-1 \\ i \text{ odd}}} (2q+2i-8n+9), \max_{\substack{1 \leq i \leq 2n-1 \\ i \text{ odd}}} (i+2n-1), \right.$$

$$\max_{\substack{2 \leq i \leq 2n-2 \\ i \text{ even}}} (2q-i-6n+5) \; i+2j-2m-2, \max_{\substack{1 \leq i \leq 2n-1 \\ i \text{ odd} \\ 1 \leq j \leq m}} (2m+1)\,i+2j-2m-2, \max_{\substack{2 \leq i \leq 2n-2 \\ i \text{ even} \\ 1 \leq j \leq m}} 2q-(2m+1)\,i-2j+2m+2,$$

$$\max_{\substack{1 \leq i \leq 2n-1 \\ i \text{ odd} \\ 1 \leq j \leq m}} 2q+(2m+4)\,i+2j-(6m+6)\,n, \max_{\substack{1 \leq i \leq 2n-1 \\ i \text{ odd} \\ 1 \leq j \leq m}} 2q+(2m+1)\,i+2j-(4m+10)\,n+6,$$

$$\left. \max_{\substack{1 \leq i \leq 2n-1 \\ i \text{ even} \\ 1 \leq j \leq m}} q-(2m+1)\,i-2j-m\,n+2m+2 \right\} = 2q-1$$

, The maximum value of all odds. Thus $\phi(v) \in \{0, 1, 2, \ldots, 2q-1\}$

(b) Clearly $\phi$ is a one – to – one mapping from the vertex set of $G$ to $\{0, 1, 2, \ldots, 2q-1\}$.

c) It remains to show that the labels of the edges of $G$ are all the odds of the interval $[1, 2q-1]$.

The range of $|\phi(v_j) - \phi(v_i)| = \{2q - j - i + 2 \quad , j\text{-even, } i\text{-odd}\}$

The range of $|\phi(u_j) - \phi(u_i)| = \{2q - j - i - 8n + 6 \quad , j\text{-even, } i\text{-odd}\}$

The range of $|\phi(u_j) - \phi(w_i)| = \{2q - 6n + 3i - 4 \quad , j\text{-odd}, i = 1, 2, \ldots n\}$

The range of $|\phi(v_j) - \phi(w_i)| = \{2q - 4n + 3i - 4 \quad i\text{ – odd}, 1 \leq i \leq 2n - 1\}$

The range of $|\phi(v_i^j) - \phi(v_i)| = \{2m\,i + 2j - 2m - 1 \; , i\text{ –odd}, 1 \leq j \leq m\}$

The range of $|\phi(v_i^j) - \phi(v_i)| = \{2m\,i + 2j - 2m - 1 \; , i\text{ –even}, 1 \leq j \leq m\}$

The range of $|\phi(w_i^j) - \phi(w_i)| = \{6n(m+1) - 2mi - 2j - 4n + 5 \; ; i\text{ – odd}, 1 \leq i \leq 2n-1, 1 \leq j \leq m\}$

The range of $|\phi(u_i^j) - \phi(u_i)| = \{2q + 2m\,i + 2j - (4m+8)n + 7 \; , i\text{ –odd}, 1 \leq j \leq m\}$

The range of $|\phi(u_i^j) - \phi(u_i)| = \{q + 2m\,i + 2j + (m-6)n - 2m + 3 \; , i\text{ –even}, 1 \leq j \leq m\}$

Hence $\{|\phi(u) - \phi(v)|: uv \in E\} = \{1, 3, \ldots, 2q-1\}$ so that graph $S(L_n) \odot mk_1$ is odd graceful.

∎

Rosa [1] defined a triangular snake (or $\Delta$-snake) as a connected graph in which all blocks are triangles and the block-cut-point graph is a path. Let $\Delta_k$-snake be a $\Delta$-snake with $k$ blocks while $n\Delta_k$-snake be a $\Delta$-snake with $k$ blocks and every block has $n$ number of triangles with one common edge. Badr and Abdel-aal [12] proved that an odd graceful labeling of the all subdivision of double triangular snakes ($2\Delta_k$-snake).

**Theorem3:** All the subdivision of triangular snakes ($\Delta_k - snake$) with *m-pendant* edges $S(\Delta_k - snake) \odot mk_1$ are odd graceful.

**Proof:**

Let $G = S(\Delta_k - snake) \odot mk_1$ has $q$ edges and $p$ vertices. The graph $\Delta_k - snake$ consists of the vertices $u_i$ and $w_j$ where $i = 1, 2, \ldots, k+1$ and $j = 1, 2, \ldots, k$. In order to get the subdivision of the $\Delta_k - snake$, we run the following steps:

i) Let $y_i$ be the newly added vertex between $u_i$ and $u_{i+1}$ where $i = 1, 2, \ldots, k$.
ii) Let $v_i$ be the newly added vertex between $u_i$ and $w_i$ where $i = 1, 2, \ldots, k$.
iii) Let $z_i$ be the newly added vertex between $w_i$ and $u_{i+1}$ where $i = 1, 2, \ldots, k$.

iv) Now, in order to get the subdivision of triangular snakes $S(\Delta_k - snake)$ with pendant edges, for each $u_i, w_j, y_j, v_j$ and $z_j$ connect with m-pendant edges $u_i^j, w_i^j, v_i^j, v_i^j$ and $z_i^j$ where $i = 1, 2, …, k+1$, $j = 1, 2, …, k$ and $l = 1, 2, …, m$ as shown in Figure 1.

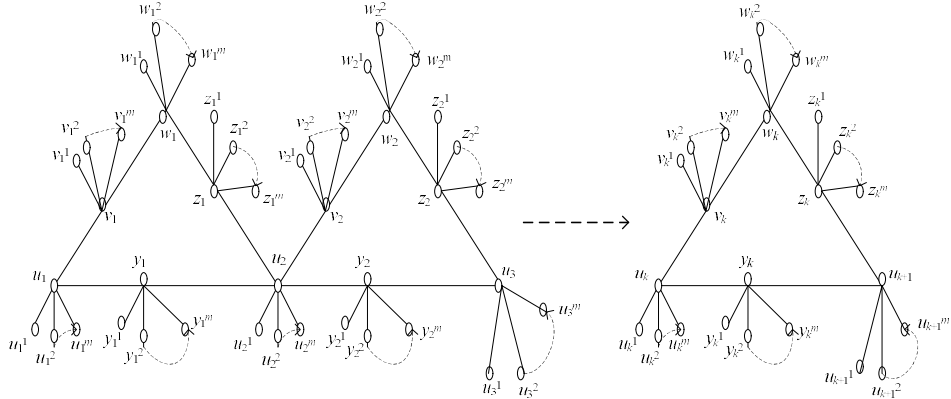

Figure 1: The graph $S(\Delta_k - snake) \odot mk_1$

Clearly, the graph $G = (m, k) \Delta_k - snake$ has $q = (5m+6)k + m$ edges and $p = (5k+1)(m+1)$ vertices. We prove that all the subdivision of triangular snakes ($\Delta_k - snake$) with pendant edges $S(\Delta_k - snake) \odot mk_1$ are odd graceful. Let us consider the following numbering $\phi$ of the vertices of the graph $G$:

$\phi(u_i) = (4m + 4)(i - 1)$     $i = 1, 2, 3… k+1$.
$\phi(w_i) = (4m + 4) i - 2m - 2$     $i = 1, 2, 3… k$.
$\phi(v_i) = 2q - (2m + 4) i + 2m + 3$     $i = 1, 2, 3… k$.
$\phi(z_i) = 2q - (2m +4) i +1$     $i = 1, 2, 3… k$.
$\phi(y_i) = (4m +4) k – (4m + 4) i + 4m + 3$     $i = 1, 2, 3… k$.
$\phi(w_i^l) = 2q – (2m + 4) i – 2l + 2m +3$     $, i = 1, 2, 3… k$ $, l = 1, 2… m$.
$\phi(v_i^l) = (4m + 4) i + 2l - 4m - 4$     $, i = 1, 2, 3… k$ $, l = 1, 2… m$.
$\phi(z_i^l) = (4m + 4) i + 2l - 2m - 2$     $, i = 1, 2, 3… k$ $, l = 1, 2… m$.
$\phi(u_1^l) = 2l + 1$     $, l = 1, 2… m$.
$\phi(u_{k+1}^l) = 2q – 2l – (2m +4)(k+1) + 2m + 5$     $, l = 1, 2… m$
$\phi(y_1^l) = 2q – 2l – (2m + 4) k - 2m - 2$     $, l = 1, 2… m$

If $k - even$ $(k = 2, 4, 6….)$

$f(u_i^l) = \begin{cases} q + (2m + 2) i - 2l - (3m + 4) k - m - 1 & : i, k\text{-even}; 2 \le i \le k; 1 \le l \le m \\ q + (2m + 2) i - 2l - (3m + 4) k - m + 1 & : k\text{-even}; i\text{-odd } 2 \le i \le k-1; 1 \le l \le m \end{cases}$

$f(v_i^l) = \begin{cases} q - (2m + 2) i + (k + 1) m - 2l + 2 & : i, k\text{-even}; 2 \le i \le k-2; 1 \le l \le m \\ q - (2m + 2) i + (k + 1) m - 2l & : k\text{-even}; i\text{-odd } 2 \le i \le k-1; 1 \le l \le m \end{cases}$

If $k - odd$ $(k = 1, 3, 5…)$

$f(u_i^l) = \begin{cases} q + (2m + 2) i - 2l - (3m + 4) k - m + 1 & : k\text{-odd}, i\text{-even}; 2 \le i \le k-1; 1 \le l \le m \\ q + (2m + 2) i - 2l - (3m + 4) k - m - 1 & : i, k\text{-odd}; 2 \le i \le k; 1 \le l \le m \end{cases}$

$f(v_i^l) = \begin{cases} q - (2m + 2) i + (k + 1) m - 2l & : k\text{-odd}; i\text{-even}; 2 \le i \le k-1; 1 \le l \le m \\ q - (2m + 2) i + (k + 1) m - 2l + 2 & : i, k\text{-odd}; 2 \le i \le k-3; 1 \le l \le m \end{cases}$

(a)
$$Max_{v \in V}\phi(v) = \max \left\{ \max_{1 \leq i \leq k+1}(4m+4)(i-1), \max_{1 \leq i \leq k}(4m+4)i - 2m - 2, \max_{1 \leq i \leq k}2q - (2m+4)i + 2m + 3, \max_{1 \leq i \leq k}2q - (2m+4)i + 1, \right.$$

$$\max_{\substack{1 \leq i \leq k \\ 1 \leq l \leq m}}(4m+4)k - (4m+4)i + 4m + 3, \max_{\substack{1 \leq i \leq k \\ 1 \leq l \leq m}}2q - (2m+4)i - 2l + 2m + 3, \max_{\substack{1 \leq i \leq k \\ 1 \leq l \leq m}}(4m+4)i + 2l - 4m - 4,$$

$$\max_{\substack{1 \leq i \leq k \\ 1 \leq l \leq m}}(4m+4)i + 2l - 2m - 2, \max_{1 \leq l \leq m}2l + 1, \max_{1 \leq l \leq m}2q - 2l - (2m+4)(k+1) + 2m + 5, \max_{1 \leq l \leq m}2q - 2l - (2m+4)k - 2m - 2,$$

$$\max_{\substack{2 \leq i \leq k \\ 1 \leq l \leq m}}^{i,k\,even} q + (2m+2)i - 2l - (3m+4)k - m - 1, \max_{\substack{2 \leq i \leq k-1 \\ 1 \leq l \leq m}}^{k\,even, i\,odd} q + (2m+2)i - 2l - (3m+4)k - m + 1,$$

$$\max_{\substack{2 \leq i \leq k-2 \\ 1 \leq l \leq m}}^{i,k\,even} q - (2m+2)i + (k+1)m - 2l + 2, \max_{\substack{2 \leq i \leq k-1 \\ 1 \leq l \leq m}}^{k\,even, i\,odd} q - (2m+2)i + (k+1)m - 2l,$$

$$\max_{\substack{2 \leq i \leq k-1 \\ 1 \leq l \leq m}}^{i\,even, k\,odd} q + (2m+2)i - 2l - (3m+4)k - m + 1, \max_{\substack{2 \leq i \leq k \\ 1 \leq l \leq m}}^{i,k\,odd} q + (2m+2)i - 2l - (3m+4)k - m - 1,$$

$$\left. \max_{\substack{2 \leq i \leq k-1 \\ 1 \leq l \leq m}}^{i\,even, k\,odd} q - (2m+2)i + (k+1)m - 2l, \max_{\substack{2 \leq i \leq k-3 \\ 1 \leq l \leq m}}^{i,k\,odd} q - (2m+2)i + (k+1)m - 2l + 2 \right\} = 2q - 1$$

,the maximum value of all odd integers. Thus $\phi(v) \in \{0, 1, 2 \ldots, 2q-1\}$
(b) Clearly $\phi$ is a one – to – one mapping from the vertex set of $G$ to $\{0, 1, 2, \ldots, 2q-1\}$.
(c) It remains to show that the labels of the edges of $G$ are all the odds of the interval $[1, 2q-1]$.

The range of $|\phi(v_i) - \phi(u_i)| = \{2q - (6m + 8)i + 6m + 7$ , $i = 1, 2 \ldots k\}$
The range of $|\phi(v_i) - \phi(w_i)| = \{2q - (6m + 8)i + 4m + 5$ , $i = 1, 2 \ldots k\}$
The range of $|\phi(z_i) - \phi(w_i)| = \{2q - (6m + 8)i + 2m + 3$ , $i = 1, 2 \ldots k\}$
The range of $|\phi(z_i) - \phi(u_{i+1})| = \{2q - (6m + 8)i + 1$ , $i = 1, 2 \ldots k\}$
The range of $|\phi(y_i) - \phi(u_i)| = \{(4m + 4)k - (8m + 8)i + 8m + 7$ , $i = 1, 2 \ldots k\}$
The range of $|\phi(y_i) - \phi(u_{i+1})| = \{(4m + 4)k - (8m + 8)i + 4m + 3$ , $i = 1, 2 \ldots k\}$
The range of $|\phi(u_1^l) - \phi(u_1)| = \{2l + 1$ , $l = 1, 2 \ldots m\}$
The range of $|\phi(u_p^l) - \phi(u_p)| = \{2q - (6m + 8)p - 2l + 6m + 9$ , $l = 1, 2 \ldots m\}$
The range of $|\phi(y_k^l) - \phi(y_k)| = \{2q - (2m + 4)k - 2l - 6m - 5$ , $i = 1, 2 \ldots k, l = 1, 2 \ldots m\}$
The range of $|\phi(w_i^l) - \phi(w_i)| = \{2q - (6m + 8)i - 2l + 4m + 5$ , $i = 1, 2 \ldots k, l = 1, 2 \ldots m\}$
The range of $|\phi(v_i^l) - \phi(v_i)| = \{2q - (6m + 8)i - 2l + 6m + 7$ , $i = 1, 2 \ldots k, l = 1, 2 \ldots m\}$
The range of $|\phi(z_i^l) - \phi(z_i)| = \{2q - (6m + 8)i - 2l + 2m + 3$ , $i = 1, 2 \ldots k, l = 1, 2 \ldots m\}$
The range of $|\phi(u_i^l) - \phi(u_i)| = \{q - (2m + 2)i - (3m + 4)k - 2l + 3m + 3$ , $k$-even, $i$-even $1 \leq l \leq m\}$
The range of $|\phi(u_i^l) - \phi(u_i)| = \{q - (2m + 2)i - (3m + 4)k - 2l + 3m + 5$ , $k$-even, $i$-odd, $1 \leq l \leq m\}$
The range of $|\phi(u_i^l) - \phi(u_i)| = \{q - (2m + 2)i - (3m + 4)k - 2l + 3m + 5$ , $k$-odd, $i$-even, $1 \leq l \leq m\}$
The range of $|\phi(u_i^l) - \phi(u_i)| = \{q - (2m + 2)i - (3m + 4)k - 2l + 3m + 3$ , $k$-even, $i$-odd $1 \leq l \leq m\}$
The range of $|\phi(y_i^l) - \phi(y_i)| = \{q + (2m + 2)i - (3m + 4)k - 2l - 3m - 1$ , $k$-even, $i$-even $1 \leq l \leq m\}$
The range of $|\phi(y_i^l) - \phi(y_i)| = \{q + (2m + 2)i - (3m + 4)k - 2l - 3m - 3$ , $k$-even, $i$-odd, $1 \leq l \leq m\}$

The range of $|\phi(y_i^l) - \phi(y_i)| = \{q + (2m + 2)i - (3m + 4)k - 2l - 3m - 3$ , $k$-odd, $i$-odd, $1 \leq l \leq m\}$

The range of $|\phi(y_i^l) - \phi(y_i)| = \{q + (2m + 2)i - (3m + 4)k - 2l - 3m - 1$ , $k$-odd, $i$-even, $1 \leq l \leq m\}$

Hence $\{|\phi(u) - \phi(v)|: uv \in E\} = \{1, 3, \ldots, 2q-1\}$ so that graph $G = S(\Delta_k - snake) \odot mk_1$ is odd graceful.

∎

**Example 4**

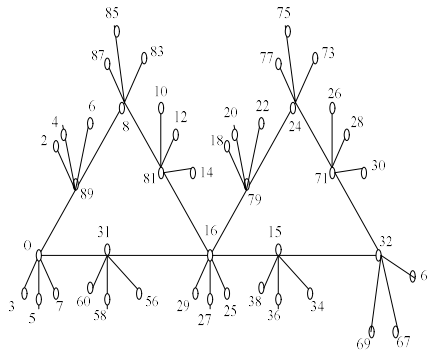

Figure 2: the odd graceful labeling of The graph S ($2C_3$-snake) ⊙ $2k_1$-snake

## CONCLUSION:

In this paper, we show that the ladder graph $L_n$ with $m$-pendant $L_n \odot mk_1$ is odd graceful. We also show that the subdivision of ladder graph $L_n$ with $m$-pendant $S(L_n) \odot mk_1$ is odd graceful. Finally, we proved that all the subdivision of triangular snakes ($\Delta_k - snake$) with pendant edges $S(\Delta_k - snake) \odot mk_1$ are odd graceful.